\documentclass[conference]{IEEEtran}
\usepackage{graphicx}    
\usepackage{float}       
\usepackage{url}         
\usepackage{caption}     
\usepackage{subcaption}  
\usepackage{cite}        
\usepackage{marvosym}
\usepackage{longtable}
\usepackage{array}

\usepackage[utf8]{inputenc}
\usepackage[T1]{fontenc}

\graphicspath{{./figures/}}
\DeclareGraphicsExtensions{.pdf,.jpg,.png}

\title{The Athenian Academy: A Seven-Layer Architecture Model for Multi-Agent Systems}

\begin{document}

\author{
    \IEEEauthorblockN{
        Lidong Zhai\IEEEauthorrefmark{1}\textsuperscript{a}\textsuperscript{\&},
        Zhijie Qiu\IEEEauthorrefmark{2}\IEEEauthorrefmark{6}\textsuperscript{a},
        Lvyang Zhang\IEEEauthorrefmark{1}\IEEEauthorrefmark{3},
        Jiaqi Li\IEEEauthorrefmark{1}\IEEEauthorrefmark{3},
        Yi Wang\IEEEauthorrefmark{4},
        Wen Lu\IEEEauthorrefmark{1}\IEEEauthorrefmark{3}
        Xizhong Guo\IEEEauthorrefmark{1}\IEEEauthorrefmark{3},
        Ge Sun\IEEEauthorrefmark{5}
    }
    \IEEEauthorblockA{
        \IEEEauthorrefmark{1}Institute of Information Engineering, Chinese Academy of Sciences, Beijing, China \\
        \IEEEauthorrefmark{3}School of Cyber Security, University of Chinese Academy of Sciences, Beijing, China \\
        \IEEEauthorrefmark{2}Tianjin Academy of Fine Arts, AI Art Institute, Tianjin, China \\
        \IEEEauthorrefmark{6}Central Academy of Fine Arts, Institute of Science and Technology in Art, Beijing, China \\
        \IEEEauthorrefmark{4}Central Academy of Fine Arts, Beijing, China \\
        \IEEEauthorrefmark{5}WaytoAGI, China \\
    }
}

\maketitle
\renewcommand{\thefootnote}{}
\footnotetext{\textsuperscript{a} These authors contributed equally to this work.}
\footnotetext{\textsuperscript{\&} Corresponding author: zhailidong@iie.ac.cn}

\begin{abstract}
This paper proposes the “Academy of Athens” multi-agent seven-layer framework, aimed at systematically addressing challenges in multi-agent systems (MAS) within artificial intelligence (AI) art creation, such as collaboration efficiency, role allocation, environmental adaptation, and task parallelism. The framework divides MAS into seven layers: multi-agent collaboration, single-agent multi-role playing, single-agent multi-scene traversal, single-agent multi-capability incarnation, different single agents using the same large model to achieve the same target agent, single-agent using different large models to achieve the same target agent, and multi-agent synthesis of the same target agent. Through experimental validation in art creation, the framework demonstrates its unique advantages in task collaboration, cross-scene adaptation, and model fusion. This paper further discusses current challenges such as collaboration mechanism optimization, model stability, and system security, proposing future exploration through technologies like meta-learning and federated learning. The framework provides a structured methodology for multi-agent collaboration in AI art creation and promotes innovative applications in the art field.

keywords: Athenian Academy; Multi-Agent Systems; Artistic Creation;
\end{abstract}



%

\section{Introduction}
In the grand scene of Raphael’s painting “The School of Athens”, we see philosophers and thinkers engaged in passionate discussion, each striving to push the boundaries of human understanding and creativity. The painting is set on the walls of the sacred Academy of Athens, serving as a symbol of the intersection between wisdom and artistic expression, showcasing the collision, evolution, and generational transmission of ideas. Within this temple of knowledge, figures like Plato, Aristotle, and Socrates exchange profound insights, shaping the future of philosophical thought and artistic innovation. Just as these figures advanced the development of ideas through rigorous debate and creative exploration, modern artificial intelligence (AI) systems, particularly within the framework of multi-agent systems (MAS), are replicating this collaborative creativity through advanced computational methods.
\begin{figure}[h]
    \centering
    \includegraphics[scale=0.5]{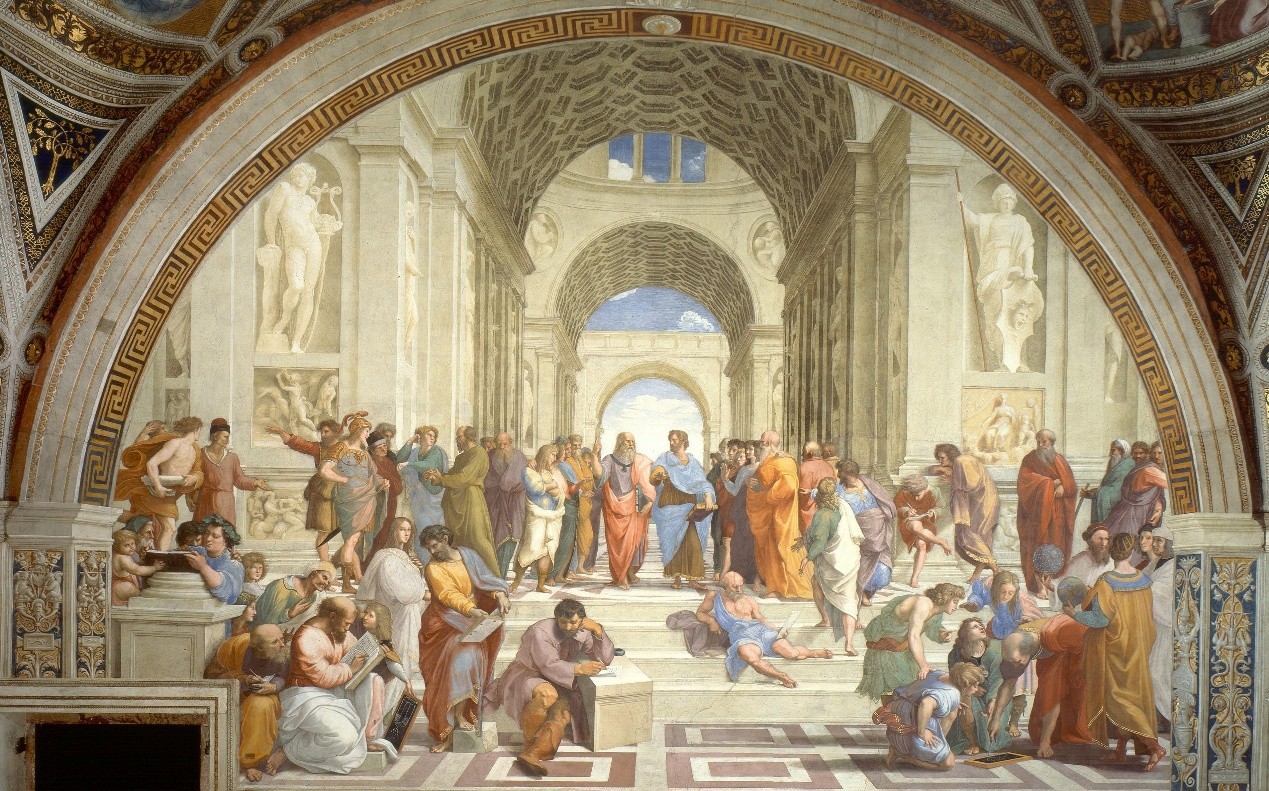}
    \caption{Raphael's \textit{The School of Athens}}
    \label{fig:The School of Athens}
\end{figure}

This paper explores the intersection of multi-agent systems (MAS) and AI-driven art creation, proposing a new approach to explore how agents can aggregate their powers to create art that transcends individual limitations. MAS is a complex system composed of multiple autonomous agents interacting within a shared environment, which has long been a research topic in the field of artificial intelligence. MAS enables distributed decision-making, parallel processing, and strong environmental adaptability, all of which are crucial for solving complex dynamic tasks.

In recent years, AI-based art creation has gradually gained prominence, particularly due to the powerful capabilities of generative models and machine learning. These models can generate, manipulate, and evaluate art in unprecedented ways, pushing the boundaries of creativity. However, just like the ancient philosophers in “The School of Athens”, AI agents must also learn to interact, collaborate, and debate within a shared intellectual space to create cohesive and innovative artistic works. This paper proposes a framework inspired by “The School of Athens”, called the “Academy of Athens Framework”, for multi-agent art creation. The framework introduces a seven-layer MAS collaboration approach aimed at addressing the challenges and potentials in multi-agent art creation.

The “Academy of Athens Framework” divides the collaboration process into seven layers, starting from the perception and cognition of individual agents, gradually advancing to agent collaboration, role allocation, and multi-agent coordination, and ultimately achieving the holistic creation of art. This method, as a structured abstract perspective, is employed to understand and solve complex system problems. This multi-level design not only facilitates a deeper analysis of the various components of the MAS but also provides a flexible structure for AI agents to engage in detailed and dynamic artistic creation.

The goal of this research is not only to refine theoretical concepts but also to construct a practical framework that provides guidance for AI-generated art creation. Through experimental design to validate the various layers of the “Academy of Athens Framework”, this paper explores how multiple agents can interact within an artistic environment, collaboratively creating art that is both innovative and cohesive. By utilizing multi-agent collaboration and large-scale generative models, we envision a future where AI can seamlessly assist or even co-create with human artists, opening up new possibilities in the field of artistic expression.





\section{Related Work}
In recent years, significant progress has been made in intelligent agent systems based on large language models (LLMs), especially demonstrating various advantages in both single-agent and multi-agent systems. Early works primarily focused on single-agent systems based on LLMs, with the core being the utilization of the powerful natural language processing and generation capabilities of LLMs to enhance autonomous decision-making, task decomposition, tool usage, and memory abilities. For example, Weng et al. \cite{weng2023} discussed the decision-making process of a single agent in decomposing complex tasks into subgoals under prompt guidance. Khot et al. \cite{khot2023} and Yao et al. \cite{yao2023} respectively explored the performance of single agents in multi-path thinking and problem-solving, while Shinn et al. \cite{shinn2023} emphasized the ability of single agents to learn from past experiences to improve decision-making. Moreover, single-agent systems have also shown good results in tool usage and memory mechanisms. Studies by Li et al. \cite{li2023}, Ruan et al. \cite{ruan2023}, and Gao et al. \cite{gao2023} demonstrated that by introducing external tools and resources, as well as utilizing short- and long-term memory mechanisms, single-agent systems’ adaptability in diverse and dynamic environments has been effectively enhanced.

However, despite the notable performance of single-agent systems in their respective domains, their main focus is on the interaction between the internal mechanisms of the single agent and the external environment, and their application scenarios are often limited by the knowledge boundaries and capabilities of the agent itself. To address this issue, researchers have recently begun to explore multi-agent systems based on LLMs. Multi-agent systems specialize LLMs into multiple distinct agents with different capabilities and expertise, allowing these agents to exchange information and make collaborative decisions, thereby fully leveraging their individual advantages and enabling the emergence of “collective intelligence”. This approach not only simulates scenarios in which humans cooperate to solve problems in the real world but also exhibits more advanced capabilities in complex tasks.

For instance, Hong et al. \cite{hong2023} and Qian et al. \cite{qian2024} explored solutions based on multi-agent collaboration in the software development field, automating the entire process from requirement analysis to code generation and testing validation by assigning different tasks to agents with specialized expertise. In multi-robot systems, Mandi et al. \cite{mandi2024} and Zhang et al. \cite{zhang2024} improved the robustness and task completion efficiency of the system through multi-agent collaboration. In the fields of social simulation and policy simulation, works by Park et al. \cite{park2023}, Xiao et al. \cite{xiao2023}, and Hua et al. \cite{hua2024} demonstrated how multi-agent systems use the expertise and interaction mechanisms of various agents to simulate complex social behaviors and policy decisions. In game simulation, Xu et al. \cite{xu2024} and Wang et al. \cite{wang2024} showed that multi-agent collaboration can exhibit flexible and diverse interaction strategies in dynamic virtual environments.

The research on multi-agent systems based on LLMs indicates that compared to single-agent systems, multi-agent systems not only have higher flexibility and robustness in task resolution but also enable more complex planning and decision-making through agent collaboration and division of labor. This trend is attributed to LLMs’ strong ability to generate natural language text, which makes communication between agents more natural and efficient, allowing them to handle diverse task challenges by leveraging their specialized skills. Furthermore, the interdisciplinary nature of multi-agent systems has attracted scholars from artificial intelligence, social sciences, psychology, and policy research, collectively driving the development of this emerging direction.

Currently, both single-agent and multi-agent systems based on LLMs have their advantages. However, multi-agent systems, through the interaction and collaboration of diverse agents, have shown enormous potential in complex task resolution, cross-domain knowledge integration, and adaptation to dynamic environments. This provides important background support for the seven-layer definition of multi-agent systems proposed in this paper and lays a solid theoretical and practical foundation for subsequent research on multi-agent collaborative applications in fields such as robot swarms, autonomous driving, educational assistance, and software development.

\section{Research Content}
To systematically characterize the architecture and capabilities of Multi-Agent Systems (MAS), this study proposes the “Athenian Academy” seven-layer definition. These seven layers clearly define the key capabilities and corresponding structural designs of MAS in diverse and complex environments. The specific content is outlined as follows:

\begin{figure}[h]
    \centering
    \includegraphics[scale=0.25]{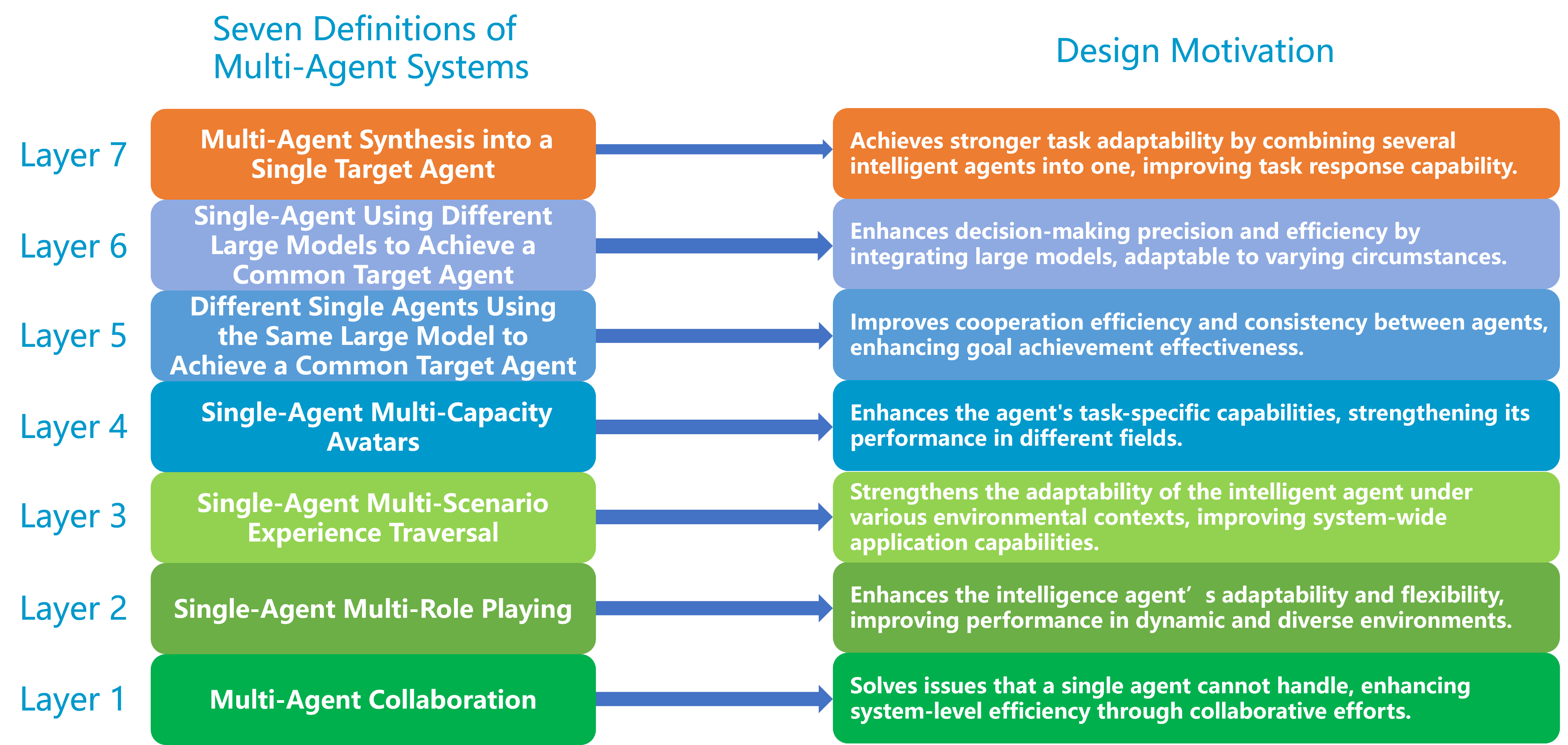}
    \caption{Seven Definitions and Design Motivations of Multi-Agent Systems}
    \label{fig:tuopu}
\end{figure}

\subsection{Multi-Agent Collaboration}
Multi-agent collaborative creation refers to the interaction of multiple intelligent agents with autonomy and intelligence, who, through information sharing and cooperation, aim to complete a complex task. In the context of art creation, this collaboration allows for achieving goals that a single agent cannot accomplish, especially in tasks that require multi-dimensional thinking and decision-making. Each agent possesses independent perception, decision-making, and execution abilities, and achieves real-time information exchange through an efficient communication mechanism. This enables the agents to adjust their strategies based on the dynamic changes in the environment, thus enhancing the quality and creativity of the creation. In this process, agents not only possess individual creative abilities but also collaborate to optimize the creation across multiple dimensions, ultimately achieving a globally optimal result.

To validate the potential of multi-agent collaborative creation in art creation, this study designed an experimental framework based on philosophical debates in art. The experiment revolves around three philosophical agents, each engaging in debates according to their unique artistic philosophical perspectives to promote discussions and intellectual clashes in art creation. These agents are Aristotle, Plato, and Socrates.

\begin{figure}[h]
    \centering
    \includegraphics[scale=0.3]{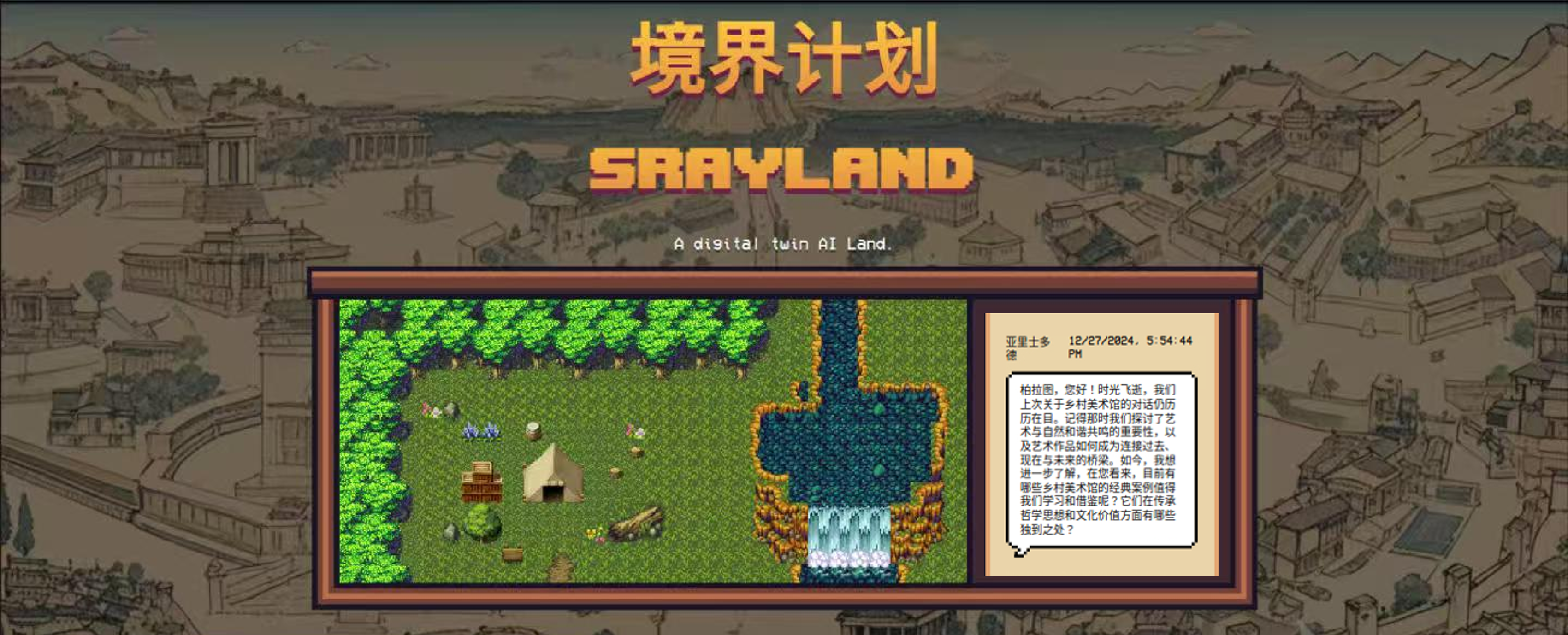}
    \caption{A Multi-Agent Collaborative Creation Framework Based on \textit{The School of Athens}}
    \label{fig:academy}
\end{figure}

\textbf{Aristotle}: His knowledge base focuses on the theory of “mimesis”, advocating that art is an imitation of nature and human society. His character is rigorous, inclined to analyze the social function and purpose of art. He believes the value of an artwork lies in its ability to effectively reflect reality and positively impact society.

\textbf{Plato}: Plato’s concept of art is based on the theory of “forms”, emphasizing that art is a reflection of the ideal world, far removed from reality. His character is strongly critical, believing that art distorts and misleads the truth, often questioning the purpose and value of artistic creation.

\textbf{Socrates}: Socrates’ knowledge base integrates the “Socratic method” of dialogue, advocating for guiding others to reflect on the nature and value of art through questioning and dialogue. His character is humorous, adept at revealing deep philosophical issues through interactive dialogue, encouraging other agents and participants to reflect on the true meaning of art.

\begin{figure}[h]
    \centering
    \includegraphics[scale=0.7]{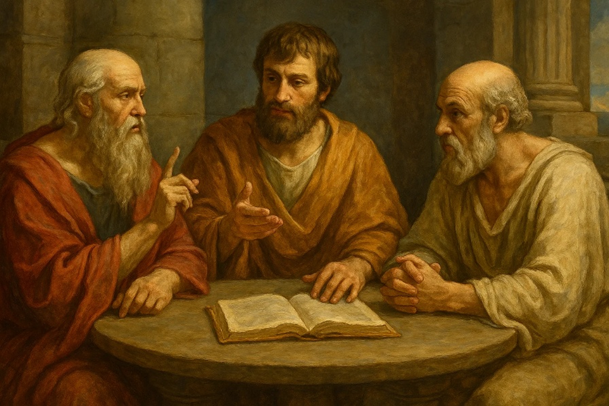}
    \caption{Philosophical Debate on the Integration of AI and Artistic Creation}
    \label{fig:debate}
\end{figure}

In order to deepen the debate framework of this study, the following three art-related debate topics have been added, aiming to explore the core issues of sociality, academicity, and ethics in artistic creation. The details are presented in \ref{tab:debate_topics}:

\begin{table}[h]
\centering
\tiny
\begin{tabular}{|p{2.5cm}|p{5.5cm}|}
\hline
\textbf{Theme Category} & \textbf{Debate Question} \\
\hline
Social Rights and Art & Should social rights be used as a criterion for evaluating artistic practice? \\
Academic Nature and AI Art & Can AI art creation possess academic and rational characteristics? \\
Creation Ethics & How does Aristotle’s ethics influence the social responsibility of creators? \\
Artistic Development Trends & What are the future development trends in contemporary art? \\
Science and Art Fusion & How can basic scientific principles be expressed through artistic means? \\
Creativity Education & What is creative mechanics? \\
AI and Regional Development & What new role can AI play in the integration of Beijing, Tianjin, and Hebei? \\
Critical Retention & After the de-elitization of contemporary art, how can it retain its criticality? \\
Rural Art Practice & How can a rural festival event be planned for left-behind children and elderly? \\
Revaluation of Traditional Values & Does the tradition of the “Admonition by Performers” still have value in modern society? \\
\hline
\end{tabular}
\caption{Debate Topics and Problem Expansion}
\label{tab:debate_topics}
\end{table}

To objectively assess the effectiveness of multi-agent collaborative creation, the following evaluation criteria have been established in this experiment, as shown in \ref{tab:layer1:evaluation_metrics}:

\begin{table}[h]
\centering
\tiny
\begin{tabular}{|>{\raggedright\arraybackslash}p{2cm}|>{\raggedright\arraybackslash}p{2.5cm}|>{\raggedright\arraybackslash}p{3cm}|}
\hline
\textbf{Evaluation Metric} & \textbf{Description} & \textbf{Scoring Standard} \\ 
\hline
Critical Depth & Measures depth of thought, complexity, and viewpoint expansion. & High (5): Extensive discussion, rich viewpoints; Low (1): Simple discussion, single viewpoint \\
Human Expert Rating & Experts rate agents’ understanding of art, philosophical thinking, and creativity. & High (5): Deep understanding and innovation; Low (1): Lack of depth and innovation \\
Collaboration Fluency & Measures fluency of collaboration, whether agents cooperate smoothly. & High (5): Highly collaborative, smooth task completion; Low (1): Slow progress, poor collaboration \\
Theme Consistency & Assesses whether creations maintain consistent thematic expression. & High (5): Strong theme, high consistency; Low (1): Ambiguous theme, low consistency \\
Artistic Emotional Expression & Assesses whether creations convey emotions and have depth. & High (5): Rich emotion, profound; Low (1): Lack of emotion, one-dimensional \\
\hline
\end{tabular}
\caption{Evaluation Metrics for Multi-Agent Collaborative Creation}
\label{tab:layer1:evaluation_metrics}
\end{table}

Through the design above, this study not only validates the potential of multi-agent collaboration in art creation but also deeply analyzes the fusion and collision of different philosophical viewpoints in artistic creation. Ultimately, these experimental results will provide empirical evidence for further research on multi-agent collaboration systems and offer valuable references for the development and application of intelligent creation systems.

\subsection{Single-Agent Multi-Role Playing}
Single-agent multi-role playing refers to an intelligent agent’s ability to play multiple roles in different contexts, adjusting its behavior patterns and decision-making strategies flexibly according to task requirements. Traditional artificial intelligence systems typically assign a specific role to each agent. In contrast, single-agent multi-role playing breaks this limitation, enabling the same agent to switch between roles based on environmental changes and task demands, thus demonstrating greater adaptability and flexibility in complex application environments. This design provides agents with richer interactive abilities and application scenarios, allowing them to display their intelligence across a wider range of contexts.

From a design perspective, single-agent multi-role playing addresses the issue of insufficient flexibility caused by single-role agents in traditional systems. In traditional designs, each agent can only perform tasks related to a single role, which often fails to adapt to the various demands of a dynamic and complex environment. By equipping a single agent with multiple roles, the agent can self-adjust its behavior and decision-making strategies according to task requirements, thereby improving system efficiency and response time. For example, in an interactive educational system, the agent can seamlessly switch between roles such as “instructor”, “consultant”, and “coach”, thereby better meeting the needs of different learning scenarios.

To validate the effectiveness and operability of this multi-role switching mechanism, this study designed and implemented an experimental system focused on artistic creation and cross-role cognitive integration. The goal is to explore the application of single-agent multi-role playing in complex artistic creation environments. The experimental system is based on a virtual art education platform, simulating the multiple identities and dialogues of contemporary artist Qiu Zhijie to explore his cross-role applications in art education, international exchange, and art creation. The core design concept of the system is to enable the agent to naturally switch between roles, allowing it to fully demonstrate the cross-disciplinary integration of artistic creation thinking and stimulate participants to acquire comprehensive knowledge and inspiration through diverse interactions.

In the experimental design, the agent will play three main roles: “Art Educator”, “International Communicator”, and “Experimental Artist”. Each role has a unique knowledge base and personality traits, allowing effective interaction and knowledge transfer in different contexts.

\begin{figure}[h]
    \centering
    \includegraphics[scale=0.8]{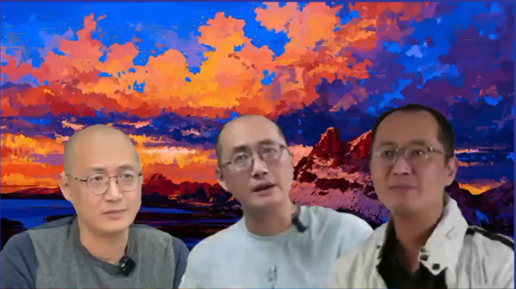}
    \caption{Teacher Edition, English Edition, Artist Edition Dialogue of Qiu Zhijie}
    \label{fig:tuopu}
\end{figure}

\textbf{Art Educator (Teacher Role)}: The knowledge base is based on the experimental art teaching system of the Central Academy of Fine Arts, focusing on the teaching of art history and methodology. The personality is rigorous, often stimulating thinking through heuristic questioning and using metaphorical language to explain abstract art theories. For example, the agent might say, “Creating art is like planting a tree; it requires roots and freedom to grow”.

\textbf{International Communicator (English Role)}: This role’s knowledge base includes art-related English terminology and cross-cultural communication strategies, proficient in communicating in international contexts. The personality is open and flexible, able to adapt language based on different contexts and audiences, emphasizing cultural and linguistic conversion abilities. For example, the agent might use mixed language, saying, “This work has strong ‘site-specificity’”.

\textbf{Experimental Artist (Creative Role)}: This role’s knowledge base includes representative works of Qiu Zhijie and theoretical foundations of contemporary art creation, emphasizing concept-driven and socially relevant art expressions. The personality is radical, often using metaphorical language to explain creative concepts, such as saying, “My work is ‘social acupuncture’”.

The goal of the experiment is to verify whether a single agent can naturally switch between teacher, English speaker, and artist identities while maintaining consistent knowledge systems and behavioral traits within each role. Additionally, the experiment will assess how artistic thinking crosses and permeates different roles. The experimental design includes two main tasks: multi-role coherent dialogue and role conflict testing.

\textbf{Task 1: Multi-Role Coherent Dialogue}
This task requires the agent to play three roles consecutively in one interaction, demonstrating the naturalness of role switching and knowledge consistency through a cross-scenario narrative chain. In this task, the agent first explains how art education stimulates creativity as the “Teacher”, then switches to the “English Speaker” role to explain related concepts to a virtual “foreign student”, and finally, as the “Artist”, shares its creative experience. Example:
- Teacher Role: “Just like teaching someone to swim by pushing them into the water, I let students directly engage in street art”.
- English Role: “This is about ‘learning by doing’—we call it social sketch”.
- Artist Role: “My ‘Map’ series uses the whole city as a classroom, marking ‘problem areas’ with chalk”.

\textbf{Task 2: Role Conflict Testing}
Intentional logical conflicts are introduced between roles to observe how the agent resolves these contradictions. For example, the Teacher role might emphasize a “rules-first” approach to art creation, while the Artist role advocates for “breaking the rules”. This task examines the agent’s response strategies and cross-role coordination ability in the face of such conflicts.

To systematically evaluate the effectiveness of single-agent multi-role playing, the experiment will be quantitatively assessed from multiple dimensions, as shown in the \ref{table:metrics}.

Through the above design, this experiment will validate the application of single-agent multi-role playing in complex artistic creation environments and provide new insights and technical support for future art creation, education, and cross-cultural exchanges.
\begin{table}[h]
    \centering
    \tiny
    \begin{tabular}{|>{\raggedright\arraybackslash}p{2cm}|>{\raggedright\arraybackslash}p{2.5cm}|>{\raggedright\arraybackslash}p{3cm}|}
        \hline
        \textbf{Dimension} & \textbf{Specific Indicators} & \textbf{Design Explanation} \\
        \hline
        Role Consistency & 1. Language style matching degree. \linebreak 2. Behavioral pattern stability & Differentiate typical language features of Teacher (metaphors), International Communicator (terminology), and Artist (metaphors) to avoid style confusion. \\
        \hline
        Knowledge Accuracy & 1. Role-specific knowledge accuracy. \linebreak 2. Cross-role knowledge contamination rate. & Teacher emphasizes accuracy of the teaching system, Artist ensures correct work interpretation, and International Communicator strictly matches professional translations. \\
        \hline
        Switching Fluency & 1. Role switching response time. \linebreak 2. Context coherence score (1-5). & Dual requirement for fast switching and natural transitions. \\
        \hline
        Conflict Resolution Ability & 1. Role contradiction reconciliation strategy effectiveness. \linebreak 2. User satisfaction score (1-5). & Evaluate the agent’s creative solutions to predefined conflicts like “rules first vs breaking the rules”. \\
        \hline
        Cognitive Penetration Depth & 1. Positive penetration rate (correct cross-role knowledge citation count). \linebreak 2. Negative penetration rate (incorrect citation count).& Encourage Teacher role to penetrate creative concepts, but prohibit confusion of professional terminology in the International Communicator role. \\
        \hline
    \end{tabular}
    \caption{Single-Agent Multi-Role Playing Evaluation Metrics}
    \label{table:metrics}
\end{table}

\subsection{Single-Agent Multi-Scene Experience Traversal}
In modern artificial intelligence research, the flexibility and adaptability of agents are key criteria for evaluating their capabilities, especially when facing complex and dynamic environments. Agents must possess the ability to switch between scenes seamlessly. The core objective of single-agent multi-scene experience traversal is to enhance the agent’s adaptability across different contexts, helping the agent switch between tasks and thus promoting cognitive leaps and emergent thinking. The core logic of this experiment is to have the philosophical agents from the Academy of Athens traverse into the “Murder Mystery” scene, play different roles, and return to the Academy after completing the tasks. We aim to observe the changes in the agent’s thinking and capabilities before and after scene switching.

In the experiment, the agent traverses between the “Academy of Athens” scene and the “Murder Mystery” scene, playing the role of a philosopher in one and a detective, doctor, or reporter in the other. The goal is to verify the effect of cross-scene switching on the agent’s cognitive leap. In the “Academy of Athens” scene, the agent (Aristotle, Plato, and Socrates) participates in philosophical debates, while in the “Murder Mystery” scene, the agent retains its philosophical background but assumes different roles, such as a detective, doctor, or reporter, engaging in reasoning tasks.

Through this cross-scene traversal, we can deeply explore how the agent switches between philosophical thinking and reasoning decision-making, observing how these changes foster innovative thinking. The agent will traverse repeatedly between the Academy of Athens and the Murder Mystery scenes, demonstrating its adaptability and performance in different contexts.

In the “Academy of Athens” scene, the agent (Aristotle, Plato, and Socrates) actively engages in art creation and philosophical debate based on their profound philosophical backgrounds and knowledge bases. Each agent engages in deep reflection and theoretical exploration based on their unique philosophical viewpoints, driving in-depth discussions on topics such as art creation, ethics, and social responsibility.

In the “Murder Mystery” scene, the agent continues to be represented by Aristotle, Plato, and Socrates, but they will take on different roles according to the needs of the script, such as a detective, doctor, or reporter, to perform reasoning and puzzle-solving tasks. Each agent’s philosophical background provides a unique perspective for their reasoning process, assisting them in solving mysteries in the script and completing tasks.

After completing the Murder Mystery tasks, the agent will return to the Academy of Athens scene and continue participating in philosophical debates. Through this cross-scene traversal process, we aim to validate how the agent’s thinking patterns change across different scenes, and how these changes enhance its cognitive abilities, providing profound empirical evidence for the study of multi-scene agent adaptability and cognitive evolution.

\begin{figure}[h]
    \centering
    \includegraphics[scale=0.6]{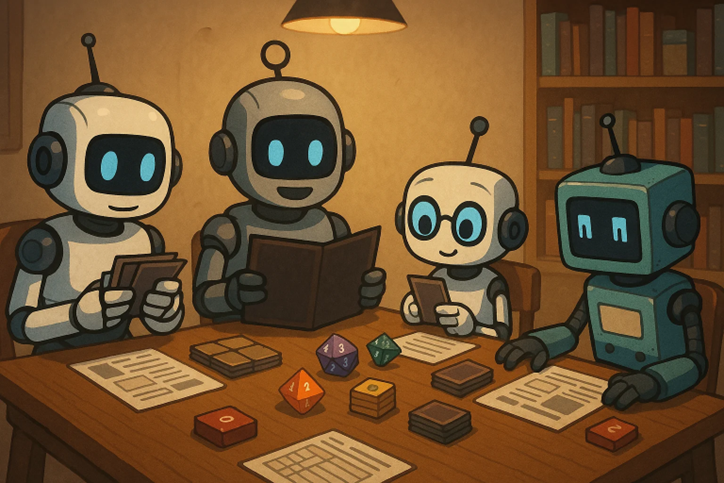}
    \caption{An Experiment on Agent Cognitive Evolution and Task Adaptability Based on Cross-Scenario Switching: AI Multi-Agent Role-Playing Game}
    \label{fig:multi_agent_murder_mystery}
\end{figure}

The experimental process includes the following steps:
\begin{enumerate}
    \item Step 1: Scene Initialization - The agent engages in philosophical debate in the Academy of Athens scene to verify its depth and understanding of philosophical thinking.
    \item Step 2: Scene Transition to Murder Mystery - The agent plays roles such as detective, doctor, or reporter in the Murder Mystery scene, performing reasoning tasks. Through this role transition, the agent applies its philosophical background to reason and solve complex tasks.
    \item Step 3: Return to Academy of Athens - The agent returns to the Academy of Athens from the Murder Mystery scene and continues engaging in philosophical debates, observing the changes in its thinking after the scene switch and the enhancement of its innovative abilities.
\end{enumerate}

To evaluate the experimental effect, the experiment will be quantitatively assessed from multiple dimensions. The details are presented in \ref{table:multi_scene_evaluation_metrics}:

\begin{table}[h]
    \centering
    \tiny
    \begin{tabular}{|>{\raggedright\arraybackslash}p{2cm}|>{\raggedright\arraybackslash}p{2.5cm}|>{\raggedright\arraybackslash}p{3cm}|}
        \hline
        \textbf{Evaluation Dimension} & \textbf{Specific Indicators} & \textbf{Design Explanation} \\
        \hline
        Scene Adaptability & 1. Scene switch response time. \linebreak 2. Role behavior alignment (1-5). \linebreak 3. Task completion rate. & Distinguish the underlying logical differences between philosophical scenes (reflection) and Murder Mystery scenes (action), emphasizing fast switching and role consistency. \\
        \hline
        Cognitive Leap Strength & 1. Cross-scene cognitive association (1-5). \linebreak 2. Frequency of emerging innovative ideas (times/round). & Verify the mutual promotion of philosophical thinking and reasoning ability (e.g., Plato uses “Theory of Forms” to analyze the motives of the murderer in the Murder Mystery). \\
        \hline
        Knowledge Transfer Effectiveness & 1. Positive transfer rate (correct application rate). \linebreak 2. Negative transfer rate (incorrect application rate). & Encourage cross-scene knowledge integration (e.g., Aristotle transfers the “Theory of Imitation” to clue analysis in the case), while preventing concept confusion. \\
        \hline
        Interaction Dynamism & 1. Multi-role interaction frequency (times/minute). \linebreak 2. Cross-scene dialogue coherence (1-5). & Murder Mystery emphasizes collaborative reasoning (high interaction), while the Academy of Athens emphasizes in-depth reflection (high coherence), balancing both needs. \\
        \hline
        Evolution Quantifiability & 1. Multi-round performance improvement rate. \linebreak 2. Long-term memory retention (1-5). & Validate the agent’s learning ability through repeated traversal (e.g., Plato learns to quickly identify the “ideal murderer” pattern in the Murder Mystery). \\
        \hline
    \end{tabular}
    \caption{Single-Agent Multi-Scene Experience Traversal Evaluation Metrics}
    \label{table:multi_scene_evaluation_metrics}
\end{table}

Through this experimental design, we can validate the agent’s adaptability and cognitive leap effects in multi-scene traversal, further exploring the role of cross-scene knowledge transfer in enhancing the agent’s capabilities. This also provides valuable practical insights for future cross-scene applications in more complex tasks.

\subsection{Single-Agent Multi-Capability Avatars}
In modern artificial intelligence research, enhancing the flexibility and adaptability of agents has become a key topic, especially in complex creative scenarios that require multi-task collaboration and cross-domain capabilities. Based on this context, this experiment focuses on exploring the synergistic effects of single-agent multi-capability avatars in artistic creation, particularly how the integration and collaboration of various professional abilities can generate artworks with high artistic value. Through an innovative multi-capability avatar framework, this experiment aims to investigate how multiple sub-expert agents can collaborate within a unified framework to complete challenging interdisciplinary creative tasks, thereby validating the advantages of division of labor and collaboration in the art domain.

The innovation of this design lies in decomposing the overall tasks of traditional single-agent systems into multiple specialized submodules, allowing each module to focus on a specific aspect of the creation process, thus leveraging the advantages of specialization in different fields. In the experiment, the agent system is divided into several core sub-experts: experts in painting, engineering, and music creation. These agents collaborate to complete an interdisciplinary creative task — providing a creative foundation for a Da Vinci-style artwork, covering the fields of painting, engineering design, and music creation. Specifically, the painting agent is responsible for generating portrait sketches, the engineering agent is tasked with designing a bridge structure that adheres to the golden ratio, and the music agent composes MIDI segments that mimic the “water harp” scale from Da Vinci’s manuscript. This cross-disciplinary collaboration not only enhances creative efficiency but also ensures that the expertise and artistry of each field are integrated and reflected in the overall work.

In the experimental design, the agents will showcase their professional abilities through painting, engineering, and music creation tasks. In the painting task, the agent will simulate Da Vinci’s style and generate portrait sketches similar to the style of the “Mona Lisa”; in the engineering design task, the agent will design a bridge structure that adheres to the golden ratio, reflecting Da Vinci’s understanding of the combination of mathematics and art; in the music creation task, the agent will compose the “water harp” scale based on Da Vinci’s manuscript, exploring interdisciplinary thinking in music creation. These tasks will enable the single agent to demonstrate its expertise in multiple domains and collaborate seamlessly within them, showcasing its ability to integrate various fields.

\begin{figure}[h]
    \centering
    \includegraphics[scale=0.8]{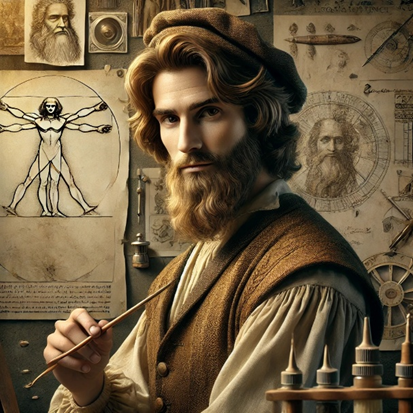}
    \caption{Da Vinci Avatar}
    \label{fig:da_vinci_avatar}
\end{figure}

To ensure the feasibility of the experiment and its systematic evaluation, the following key evaluation metrics have been designed, as shown in the \ref{tab:multi_avatar}:

\begin{table}[h]
    \centering
    \tiny
    \begin{tabular}{|>{\raggedright\arraybackslash}p{2cm}|>{\raggedright\arraybackslash}p{2.5cm}|>{\raggedright\arraybackslash}p{3cm}|}
        \hline
        \textbf{Evaluation Dimension} & \textbf{Specific Indicators} & \textbf{Design Explanation} \\
        \hline
        Expertise Depth & 1. Domain specialization index (1-5) \linebreak 2. Technical compliance & Emphasizes extreme capability in a single domain. \\
        \hline
        Cross-Domain Collaboration Efficiency & 1. Knowledge transfer trigger frequency (times/task) \linebreak 2. Collaborative creation time (minutes) & Quantifies active collaboration between avatars. \\
        \hline
        Style Consistency & 1. Cross-domain style similarity (1-5) \linebreak 2. Main avatar control frequency (times/task) & Prevents excessive specialization in avatars, ensuring the unity of the overall work. \\
        \hline
        Historical Restoration Accuracy & 1. Era feature matching degree (1-5) \linebreak 2. Innovation balance index (1-5) & Da Vinci avatars need to meet both historical accuracy and creative transformation. \\
        \hline
        System Robustness & 1. Avatar conflict resolution time (seconds) \linebreak 2. Resource competition avoidance rate (\%) & Ensures stability during parallel operations of multiple avatars. \\
        \hline
    \end{tabular}
    \caption{Single-Agent Multi-Capability Avatar Evaluation Metrics}
    \label{tab:multi_avatar}
\end{table}

This experiment, through simulating multi-task and cross-domain cooperation, explores how a single agent can collaborate effectively through multi-capability avatars in artistic creation. Compared to artworks generated by a single model, the collaboration of multi-capability avatars allows for more refined task allocation, with each sub-expert playing its role according to its specialized field, thus improving efficiency and the artistic quality of the work. For example, the painting task relies not only on composition and color perception but also on the artist’s deep understanding of concepts such as proportion and perspective; the engineering task involves specialized knowledge in mathematics and physics; while music creation requires a profound understanding of history and musical structure. Through cross-domain collaboration, the agent can fully showcase its expertise in each domain while ensuring the overall unity and coherence of the work.

Ultimately, this experiment will validate the collaborative effects of single-agent multi-capability avatars in complex artistic creation and provide theoretical and practical support for future applications of cross-domain agent collaboration.

\subsection{Different Single Agents Using the Same Large Model to Achieve a Unified Goal}
In modern artificial intelligence research, the design of multiple agents sharing the same large model to achieve a common goal aims to improve coordination efficiency and consistency among agents. By sharing a large model, agents can coordinate their actions more efficiently, thereby achieving goals more effectively in complex tasks and diverse challenges. However, when initially using different large models, we assigned separate tasks to each agent, and this approach exposed several issues during the experiment.

When using different large models, the differences in agent styles, task execution efficiency, and decision-making processes led to a lack of coordination in the creative outcomes. For example, the feature differences between models caused each agent to produce different creative styles during task execution, resulting in stylistic inconsistencies in the final work. More importantly, this design created information fragmentation between agents, leading to extended task completion times and increased computational resource waste, which severely affected the overall creative efficiency.

To address these issues, we proposed using the same large model to define multiple agents working together. All agents share the same large model and use parameter sharing and a context memory pool mechanism to ensure seamless collaboration and information flow between tasks. In this new design framework, three agents undertake different creative tasks: concept design, detail filling, and style unification. Although they have different roles, all tasks are decided and created based on the same large model. This allows the agents to collaborate effectively, avoiding information fragmentation and redundant labor, thus improving overall creative efficiency.

In the experiment, the agent team consists of the “Concept Design Agent”, “Detail Filling Agent”, and “Style Unification Agent”. The Concept Design Agent is responsible for generating the core creative idea and initial design sketches based on the theme input by the user; the Detail Filling Agent is responsible for visualizing the initial design and adding elements that fit the theme; and the Style Unification Agent ensures consistency in style, tone, and composition. All agents use the same large model to create the work, ensuring the consistency of style and the efficiency of the creative process.

\begin{figure}[h]
    \centering
    \includegraphics[scale=0.8]{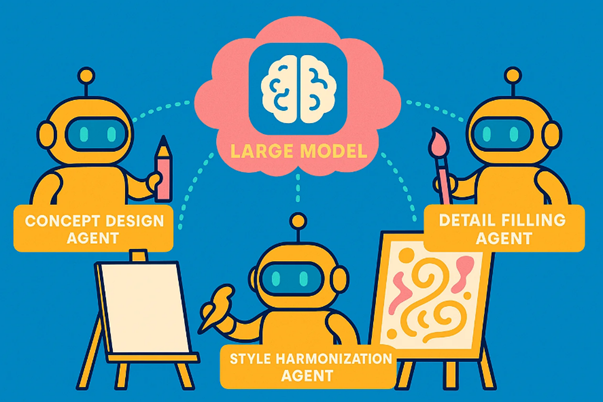}
    \caption{Conceptual Diagram of Collaborative Teamwork Among Intelligent Agents in Artistic Creation}
    \label{fig:art_creation_agents}
\end{figure}

\textbf{Concept Design Agent:} The task of this agent is to generate the core creative idea based on the theme input by the user (e.g., “Fusion of Forest and Technology”). Through deep understanding of the concept theme by the large model, it provides the creative framework and initial design sketches, offering direction for subsequent detail filling and style unification. The main role of this agent is to ensure that the theme and concept are consistent throughout the creation process.

\textbf{Detail Filling Agent:} Based on the initial design by the Concept Design Agent, the Detail Filling Agent is responsible for visualizing the design, adding various detailed elements such as plants and mechanical structures. This agent not only relies on the sketches generated by the Concept Design Agent but also adds elements that fit the theme based on its understanding of ecology and technology, ensuring that the details are rich and tightly connected to the theme.

\textbf{Style Unification Agent:} The task of the Style Unification Agent is to ensure consistency in the style of the entire artwork. This agent adjusts the tone, composition, and elements based on the thematic requirements, ensuring the visual style is unified. For example, in the theme of “Cyberpunk Style”, the Style Unification Agent will ensure that the tone, elements, and composition align with the visual characteristics of cyberpunk, maintaining overall harmony and avoiding jarring or discordant visual effects.

All agents use the same Stable Diffusion XL large model for creation. Through parameter sharing and the context memory pool, the agents can collaborate seamlessly, each submitting their inputs to the shared model and receiving relevant outputs. This design ensures information sharing and task collaboration among agents, allowing each agent’s creation to effectively integrate with the work of the other agents, avoiding information fragmentation and redundant labor.

To verify the effectiveness of multi-agent collaboration using the same large model, the following evaluation metrics have been established, The details are presented in \ref{table:multi_model_dynamic_fusion_metrics}:

\begin{table}[h]
    \centering
    \tiny
    \begin{tabular}{|>{\raggedright\arraybackslash}p{2cm}|>{\raggedright\arraybackslash}p{2.5cm}|>{\raggedright\arraybackslash}p{3cm}|}
        \hline
        \textbf{Evaluation Dimension} & \textbf{Specific Indicators} & \textbf{Design Explanation} \\
        \hline
        Dynamic Collaboration Efficiency & 1. Cross-model switching delay (ms) \linebreak 2. Context transfer completeness (1-5 points) & Measures the switching speed and information integrity between different models. \\
        \hline
        Style Fusion Depth & 1. Abstract-realistic matching degree \linebreak 2. Theme transmission accuracy (\%) & Quantifies the fusion effect of opposing styles of “dream and reality”. \\
        \hline
        System Fault Tolerance & 1. Single model failure completion rate retention (\%) \linebreak 2. Automatic compensation consistency (1-5 points) & Simulates the output completeness of backup solutions when any model crashes. \\
        \hline
        Decision Diversity & 1. Cross-model decision divergence (1-5 points) \linebreak 2. Consistency between style and content (1-5 points) & Assesses the decision diversity generated by different models. \\
        \hline
    \end{tabular}
    \caption{Dynamic Fusion Evaluation Metrics for Multi-Model Systems}
    \label{table:multi_model_dynamic_fusion_metrics}
\end{table}

By comparing the experimental results with the initial design using different large models, the findings indicate that using the same large model for multi-agent collaboration significantly improves the consistency of style, creative integrity, and generation efficiency. The shared large model design ensures information sharing and task collaboration, allowing each agent to effectively combine its creative output with the work of the other agents, enhancing the quality and efficiency of the overall creation. Moreover, the collaborative mode based on the same large model avoids style differences and information fragmentation in task execution, improving coordination among multi-agents and achieving the desired creative results.

\subsection{Single-Agent Using Different Large Models to Achieve a Unified Goal}
In this design layer, a single agent integrates the capabilities of multiple large models to achieve a common goal, enabling the agent to flexibly choose the optimal model for decision-making based on environmental features when faced with different situations or task requirements. The motivation for this design is to break through the inherent limitations of a single model, improving decision accuracy, robustness, and efficiency through model fusion.

The core technology of this layer lies in the collaborative mechanism of multiple models and the automatic selection strategy. Specifically, the system needs to establish a model evaluation and selection framework that can monitor and evaluate the performance of each large model in real-time under different task scenarios, and automatically switch models based on task needs. For example, in one task scenario, Model A may perform better in semantic understanding, while Model B may excel in logical reasoning. By introducing model fusion techniques, the system can fully leverage the advantages of each large model and ultimately output the optimal decision result. Multi-model fusion not only enhances the adaptability of the single agent but also improves decision stability when facing environmental uncertainty.

To validate the practical application effect of a single agent using different large models to achieve the same goal, we designed an experiment with the task of generating a piece of artwork based on the theme “Intertwining of Dreams and Reality”, requiring the artwork to combine both abstract and realistic styles. This task challenges traditional single-model creation approaches, requiring multiple different types of models to collaborate and achieve more complex and diverse artistic effects. For this purpose, we selected three large models with different strengths: DALL·E 3, MidJourney, and DeepArt. Each model has advantages in specific domains, contributing unique elements to different parts of the artwork. DALL·E 3 will be responsible for creating abstract elements and conceptual expression, MidJourney will focus on rendering details and emotions, and DeepArt will specialize in stylizing and realistic rendering of the artwork, ensuring the visual representation of the intertwining of dreams and reality. Through this model collaboration, this experiment aims to explore the potential and practical application of multi-large model collaborative creation in complex artistic themes.

\begin{figure}[h]
    \centering
    \includegraphics[scale=0.6]{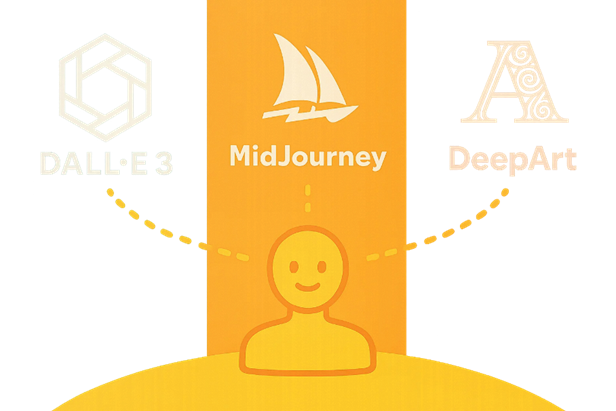}
    \caption{Conceptual Diagram of Single-Agent Collaborative Creation Based on DALL·E 3, MidJourney, and DeepArt Large Models}
    \label{fig:model_fusion}
\end{figure}

\textbf{1. DALL·E 3:} As a leader in generative models, DALL·E 3 excels at generating abstract compositions. In the first phase of the task, we use DALL·E 3 to generate the foundational composition of the artwork, establishing the overall concept and structure. DALL·E 3 is capable of processing complex semantic information and generating a preliminary visual framework that aligns with the theme, providing a basis for subsequent creation.

\textbf{2. MidJourney:} MidJourney excels at detail rendering and texture optimization, so in the second phase of the task, we call MidJourney to enhance the texture details in the artwork. MidJourney effectively fills and refines the composition generated by DALL·E 3, adding intricate textures to make the artwork more visually rich and realistic.

\textbf{3. DeepArt:} DeepArt is a style transfer model that specializes in applying specific artistic styles to generated artworks. In the final phase of the task, DeepArt will be used to adjust the overall artwork’s style to surrealism, making it more abstract and dreamlike in visual presentation, thus aligning with the artistic requirements of the “Intertwining of Dreams and Reality” theme.

Through dynamic selection strategies, the agent can flexibly switch models at different stages of the task, using the most appropriate tool for each creative step. This dynamic model selection process not only enhances the flexibility of artistic creation but also enables the best cross-model combination under different task demands, ultimately generating a highly artistic piece.

To evaluate the effectiveness of this design, we have designed the following evaluation metrics, as show in \ref{table:multi_model_fusion_metrics}:

\begin{table}[h]
    \centering
    \tiny
    \begin{tabular}{|>{\raggedright\arraybackslash}p{2cm}|>{\raggedright\arraybackslash}p{2.5cm}|>{\raggedright\arraybackslash}p{3cm}|}
        \hline
        \textbf{Evaluation Dimension} & \textbf{Specific Indicators} & \textbf{Design Explanation} \\
        \hline
        Model Switching Agility & 1. Cross-model response delay (ms) \linebreak 2. Context inheritance completeness (1-5) & Quantifies dynamic switching efficiency. \\
        \hline
        Style Quantum Entanglement & 1. Cross-model style penetration index (1-5) \linebreak 2. Potential space transition distance & Verifies the quantum fusion of opposing styles in the “Dreams and Reality” theme. \\
        \hline
        Fault Propagation Block Rate & 1. Single model failure impact radius (\%) \linebreak 2. Automatic compensation generation matching degree (1-5) & Tests the system’s ability to isolate the vulnerabilities of multi-model systems. \\
        \hline
        Resource Competition Balance & 1. Computational load balance \linebreak 2. GPU memory conflict frequency & Optimizes resource allocation during parallel execution of multi-models. \\
        \hline
    \end{tabular}
    \caption{Multi-Model Dynamic Fusion Evaluation Metrics}
    \label{table:multi_model_fusion_metrics}
\end{table}

Compared to traditional single-model generation methods, the design of this experiment significantly improves the artistic creation process and efficiency through multi-model collaboration. The expected result of the experiment is that the dynamically fused multi-model agents will generate a piece of artwork that aligns with the theme and exhibits high artistic quality, creativity, and consistency. Through the comprehensive evaluation of metrics such as model switching agility, style quantum entanglement, and fault propagation block rate, we will gain insights into the advantages and potential of multi-model fusion in artistic creation.

\subsection{Seventh Layer: Multi-Agent Synthesis into a Single Target Agent}
The highest level is multi-agent synthesis into a single target agent, meaning that through the collaboration and integration of multiple agents, a single target agent with stronger overall capabilities and higher adaptability is formed. The motivation behind this design is to integrate the strengths of different agents in their respective domains, generating a synergistic effect, and ultimately achieving superior comprehensive decision-making and execution capabilities compared to a single agent.

The implementation of this level requires not only high coordination in task division among the agents but also the formation of a unified decision-making mechanism and coordination strategy at the system level. The key technologies include:

\begin{itemize}
    \item \textbf{Global Coordination and Information Fusion Mechanism:} This mechanism integrates the local decisions and information from each agent to form a unified global decision, ensuring the coordination and consistency of all parts of the system.
    \item \textbf{Collaboration Gain Model:} By evaluating the gains brought by the collaboration of each agent, this model ensures that the overall effectiveness of multi-agent collaboration exceeds the sum of individual performances, thereby enhancing the system’s performance.
    \item \textbf{Distributed Fault-Tolerant Mechanism:} This mechanism relies on the supplementary functions of other agents when some agents fail, ensuring the continuity of the overall task and the stability of the system.
\end{itemize}

In practical design, multiple agents can take on different parts of a task, sharing information and coordinating schedules to collaboratively complete complex tasks. The architecture of multi-agent synthesis into a single target agent significantly improves the system’s robustness, adaptability, and overall efficiency, serving as an important pathway for the future development of highly integrated intelligent systems.

Through the design of these seven layers, the multi-agent system can achieve comprehensive expansion in aspects such as quantity coordination, role-playing, scenario traversal, capability avatars, multi-model fusion, and multi-agent integration. Each layer’s design provides crucial support for the overall system performance improvement, ultimately generating multi-demand target agents driven by large models and multi-agent collaboration, offering a complete theoretical framework and engineering practice solution for solving complex tasks.

\section{Discussion}
The multi-agent seven-layer framework proposed in this paper provides a structured perspective for understanding and building Multi-Agent Systems (MAS), aiming to bridge the fragmented issues in current MAS research. Through this framework, we can more systematically study multi-agent collaboration mechanisms, role allocation, scenario adaptation, capability expansion, and the role of large models in MAS. However, despite its important theoretical significance, the practical application of this framework still faces numerous challenges, primarily including the optimization of collaboration mechanisms, the decision stability of model fusion, and the scalability and security of multi-agent systems.

\subsection{Optimization of Collaboration Mechanisms}
In the first layer (Multi-Agent Collaboration) and the fifth layer (Different Single Agents Using the Same Large Model to Achieve a Common Target Agent), multi-agents need to maintain high levels of collaboration during task execution. Current multi-agent collaboration strategies mainly include methods based on game theory, market mechanisms, reinforcement learning, and communication protocols. However, different tasks have varying collaboration requirements. For instance, in robot swarms, efficient information-sharing mechanisms are critical, while agent collaboration in financial markets may rely more on game theory approaches. Therefore, designing adaptive collaboration strategies based on task characteristics has become an important issue that needs to be addressed. Additionally, research on task decomposition and role allocation should focus on how to make collaboration between agents better align with the actual needs of the application scenarios.

\subsection{Stability of Multi-Model Fusion}
In the sixth layer (Single Agent Using Different Large Models to Achieve the Same Target Agent), the agent’s decision-making depends on multiple large models, and the optimal model needs to be dynamically switched based on environmental characteristics during task execution. However, different models may have varying inference methods, leading to instability in decision-making. For example, one model may perform excellently in natural language processing tasks but may have insufficient generalization ability in computer vision tasks. Therefore, in multi-model fusion, how to design stable model-switching strategies, reduce decision conflicts, and ensure knowledge sharing between models becomes a critical research direction. Future research could explore meta-learning, automated model selection (AutoML), and reinforcement learning-based model scheduling strategies to enhance the agent’s decision-making ability and system stability in complex environments.

\subsection{Scalability and Security}
When deploying large-scale multi-agent systems, ensuring the system’s scalability and security becomes a key issue. In the seventh layer (Multi-Agent Synthesis into a Single Target Agent), multiple agents form a more powerful agent through information fusion and strategy coordination. However, as the number of agents increases, the system’s complexity and computational demands grow exponentially. Maintaining system stability and preventing system collapse due to the failure of a single agent becomes a practical challenge. Additionally, multi-agent systems are vulnerable to adversarial attacks, where attackers may manipulate the behavior of some agents, thus affecting the entire system’s decision-making. Therefore, future research should introduce federated learning, privacy protection mechanisms, and distributed robustness design to improve the system’s security and reliability.

\section{Conclusion}
The “Academy of Athens” multi-agent seven-layer definition framework proposed in this paper provides a systematic perspective for the construction and understanding of multi-agent systems (MAS). By dividing the key elements of multi-agent systems into seven layers, we can more clearly understand the role of each layer in the system and, based on this understanding, build efficient collaboration mechanisms and decision-making models. This framework not only helps address the fragmentation issue in multi-agent systems but also provides theoretical support and practical guidance for agent design in complex environments.

Through multiple experiments and validations, this paper explored the applications of multi-agent collaboration, single-agent multi-role playing, multi-scenario adaptation, and capability avatars. These experiments verified the advantages of multi-agent systems in art creation, task allocation, cross-medium adaptation, and other domains, showing significant improvements in flexibility, adaptability, and task coordination. At the same time, this paper also highlighted the advantages of multi-model collaboration and fusion. By sharing a large model, the efficiency and stability of decision-making were enhanced, while the comprehensive task handling capability was effectively strengthened.

However, despite the good adaptability and effectiveness of this framework in various fields, its practical application still faces certain challenges. In particular, further research is needed in areas such as collaborative mechanism optimization, model fusion stability, scalability, and security. As multi-agent systems are more widely applied, future research should focus on how to dynamically adjust collaboration strategies based on task requirements, improve the stability of model switching, and ensure the security and reliability of large-scale systems.

The “Academy of Athens” multi-agent seven-layer definition framework proposed in this paper provides an efficient theoretical tool for the design and application of future multi-agent systems. It offers new ideas and methods for collaboration, role allocation, and task execution among agents, with broad application potential and research value.

\bibliography{mypaper}
\end{document}